\documentclass[epsfig,graphics,twocolumn,floatfix,mathbbm,a4paper,showpacs]{revtex4}

\usepackage{amsmath,amsfonts,amssymb,graphics,graphicx,epsfig,color,times,bbm,psfrag}

\newcommand{\cc}{\mathbb{C}}
\newcommand{\nn}{\mathbb{N}}
\newcommand{\rr}{\mathbb{R}}

\newcommand{\id}{\mathbb{I}}

\newcommand{\be}{\begin{equation}}
\newcommand{\bea}{\begin{eqnarray}}
\newcommand{\eea}{\end{eqnarray}}
\newcommand{\ee}{\end{equation}}

\def\one{\ensuremath{\hbox{$\mathrm I$\kern-.6em$\mathrm 1$}}}
\def\tr{ \mbox{tr}}

\begin{document} 

\title{Half the entanglement in critical 
systems is distillable from a single specimen}

\author{R.\ Or\'us$^1$, J.I.\ Latorre$^1$,
 J.\ Eisert$^{2,3}$, and M.\ Cramer$^4$}

\affiliation{
1 Dept. d'Estructura i Constituents de la Mat\`eria,
Univ. Barcelona, 08028, Barcelona, Spain\\
2 QOLS, Blackett Laboratory, 
Imperial College London,
Prince Consort Road, London SW7 2BW, UK\\
3 Institute for Mathematical Sciences, Imperial College London,
Exhibition Rd, London SW7 2BW, UK\\
4 Institut f{\"u}r Physik, 
Universit{\"a}t Potsdam,
Am Neuen Palais 10, D-14469 Potsdam, Germany
}
\date\today

\begin{abstract}
We establish a quantitative relationship between
the entanglement content of 
a single quantum chain
at a critical point and the respective 
entropy of entanglement. 
We find that surprisingly, the leading critical scaling
of the single-copy entanglement 
with respect to any bi-partitioning 
is exactly {\it one half} of the entropy of entanglement,  in a 
general setting of conformal field theory and quasi-free
systems.
Conformal symmetry imposes that the single-copy 
entanglement  scales as
$E_1(\rho_L)=(c/6) \log L- (c/6) (\pi^2/\log L)
+ O(1/L)$, where $L$ is the number of constituents 
in a block of an infinite chain and 
$c$ denotes the central charge. 
This shows that from a single specimen of a 
critical chain, already half the entanglement 
can be distilled compared
to the rate that is asymptotically available. The 
result is substantiated by a quantitative 
analysis for all translationally invariant
quantum spin chains corresponding to
all isotropic quasi-free fermionic models. 
An example of the XY
spin chain shows that away from criticality the above
relation is only maintained near the quantum phase transition. 
\end{abstract}


\maketitle

How much entanglement is contained in a 
many-body system at zero temperature?
Variants of this question have 
received a significant attention in recent years,
notably in the context of condensed matter systems 
\cite{Osterloh,Harmonic,Latorre1,Fannes,Cardy,Peschel,LatorreNew,GraphStates,Korepin1,Area,Korepin2,Wolf,Orus,Orus2,Verstraete,Singleshot}.
In particular, 
it has turned out  that the scaling
of entanglement quantities -- similar to that of
two-point correlators \cite{Sachdev} -- 
is indeed intimately 
intertwined with critical behavior, and that signatures of
quantum phase transitions become manifest. In one-dimensional
systems in particular, it has been found that criticality
is typically accompanied with the entanglement of a 
subblock consisting of a number of consecutive 
constituents to be logarithmically divergent \cite{Latorre1,Fannes,Cardy}. 
Such a  behavior of the entropy of a subblock has also been
linked to the performance of  numerical 
DMRG-type simulations in many-body systems
close to critical points \cite{Latorre1,Verstraete}.
This quantity has a clearcut interpretation in entanglement
theory: the entropy measures the degree of entanglement,
in that it determines the optimal rate at which maximally
entangled pairs can be distilled
from a given state.
Such a procedure may invoke any collective
local quantum operations, assisted with classical 
communication (LOCC),
under the assumption that one has infinitely many identically
prepared spin systems at hand \cite{Bennett}. 
So in the present context, it
would quantify the entanglement in this asymptotic
sense, when operating locally 
on a subblock and the rest of the
system, but on many identical systems.

Needless to say, one may equally reasonably 
ask: how much entanglement
is contained in a single specimen of a many-body system?
This is meant as the largest entanglement content that any
apparatus could potentially distill with certainty 
from just one 
quantum chain at hand, resembling the situation that one
would actually encounter in any experiment. 
More specifically: what is the largest dimension of a maximally
entangled state -- or, equivalently, the maximum number of
maximally entangled qubit pairs -- 
that can be distilled with certainty from a single specimen
of a system with any physical device? The logarithm of this
quantity, introduced 
in Ref.\ \cite{Singleshot}, 
will be referred to as single-copy entanglement. That is,
for a state $\rho$ of a one-dimensional chain
with reduction $\rho_L$ to a block consisting
of $L$ consecutive constituents we write for the single-copy
entanglement  $E_1(\rho_L) = \log  M $
if $\rho\longmapsto |\psi_M\rangle\langle\psi_M|$
under LOCC, where $|\psi_M\rangle = M^{-1/2} \sum_{i=1}^M 
|i,i\rangle$ \cite{Basis}. 
Noting that single-copy transformation of pure states
under LOCC is governed by a majorization relation 
to the reduced states \cite{Nielsen}, one finds that 
$E_1(\rho_L) = \log \lfloor 1/\lambda_1 \rfloor$, where
$\lambda_1$ is the largest eigenvalue of the reduced state
$\rho_L$ of a block of length $L$ \cite{Prob}. 

In this work, we establish a fully quantitative relationship
between this single-copy quantity and the geometric
entropy -- the entropy of entanglement -- 
valid in a very large class of many-body systems
at a critical point. More specifically, for a
subblock of length $L$ we compare the single-copy
entanglement $E_1(\rho_L)$  
with the entropy of entanglement $S(\rho_L)=- 
\tr[\rho_L
\log \rho_L]$ 
\cite{loga}.
We invoke the machinery
of conformal field theory \cite{Ginsparg}
and of quasi-free systems
to relate these entanglement
contents for a single specimen and the maximal
asymptotically achievable rate. 
Conformal symmetry will
reveal a result that would otherwise appear mysterious:
we find in this setting of conformal field theory 
that the single-copy entanglement is just half the entropy of entanglement, in the
leading contribution, i.e., 
\begin{equation}
	\lim_{L\rightarrow\infty} \frac{S(\rho_L)}{E_1(\rho_L)}=2.
\end{equation}
In a {\it single run}, with a single invokation of a physical device
acting on one system, one can obtain half the
entanglement per specimen that is asymptotically available.
This also gives a guideline how much entanglement one can
expect to observe in actual single specimens of critical
quantum systems.

This result also reveals an intriguing relationship 
between the largest eigenvalue of the reduction $\rho_L$
and its full spectrum of the reduction in a very large class
of critical systems in the context of conformal field theory. 
These findings will be further substantiated by 
analogous results on a chain: for all translationally invariant 
quantum spin Hamiltonians that can be mapped onto
isotropic quadratic fermionic Hamiltonians under Jordan-Wigner
transformations \cite{Sachdev}, 
we find that if the entropy of a block is
logarithmically divergent, so is the single-copy entanglement,
with a factor of two difference in the prefactor. We finally check with the
analytical example of the XY spin chain that, away from criticality, this simple 
relation between single-copy entanglement and entanglement
entropy only holds close to the quantum phase transition point. 
 
{\it Exact conformal field theory computation. --} 
Physical properties of
quantum many-body systems are dictated at 
criticality by the
underlying symmetry under scale transformations. 
If these systems 
are described by 
means of a quantum field theory setting, it can be
 seen that the underlying symmetry group 
is even larger and becomes the so-called conformal group of
transformations. Many body systems on a lattice such as
spin chains at a critical point are assessible by a 
conformal field theory that is invariant under 
the conformal group. 
In $1$ spatial dimension, this group completely determines
the physics of the system at hand \cite{Ginsparg}. 
A key role is played by the  
central charge $c$ of the system, 
the value of which will depend on the particular
theory under consideration. Our result
particularizes to a wide variety of quantum 
chains at criticality, such as the quantum XX spin
model ($c=1$), the critical quantum Ising model
($c=1/2$) or the critical $3$-state Potts model ($c=4/5$) \cite{Ginsparg}. In our setting, for a block of size $L$,
conformal field theory provides us with an expression for the
reduced density matrix, which describes 
the vacuum of the theory. We find \cite{Orus2,Holzhey, Ginsparg}
\be 
	\label{rho}
	\rho_L=\frac{1}{Z_L(q)} q^{-c/12} 
	q^{(L_0 + \bar L_0)} \ ,
\ee
where $L_0$ and $\bar{L}_0$ are some positive semi-definite operators, 
$Z_L(q)=q^{-c/12} {\text{tr}} [q^{(L_0+\bar L_0)} ]$ is the partition 
function on a torus, 
$q=e^{2 \pi i \tau}$,  $\tau= (i \pi)/ (\log (L/\epsilon))$ 
being the so-called modular parameter, $c$ is the central charge, 
and $\epsilon$ being a short-distance cut-off which regularizes the theory. For the particularly important case of
critical quantum chains $\epsilon = 1$, 
corresponding to the lattice
spacing of the chain \cite{cutoff}

The largest eigenvalue of the density 
matrix corresponds to the
zero eigenvalue of $(L_0+\bar L_0)$, that is,
\be
	\label{largest}
	\lambda_1= \frac{1}{Z_L(q)} q^{-c/12} .
\ee
We then get for the single-copy entanglement $
	E_1(\rho_L)= \log \lfloor
	1/\lambda_1 \rfloor =  \log  \lfloor Z_L(q) q^{c/12} \rfloor $.
The leading behavior for the partition
function can be computed 
when  $L$ is large by taking advantage of its invariance
under the so-called modular transformations \cite{Ginsparg,modular}.
It is now possible to expand the partition function in powers
of $\tilde q$, $\tilde q=e^{-2\log L}$, 
as being done in Refs.\ \cite{Orus2,Holzhey},
and find that the leading contribution originates 
from the central charge $c$, 
$\log Z_L(\tilde q) = -(c/12) \log \tilde
	q+O\left(1/L\right) = (c/6)\log L+O\left(1/L\right)$.
This result translates into an explicit expression for the
single-copy entanglement
\be
\label{singlecopyL}
	E_1(\rho_L)=\frac{c}{6}\log L - \frac{c}{6} \frac{\pi^2}{\log L}+
	O  \left(1/L\right).
\ee
Eq. (\ref{singlecopyL}) is exact up to polynomial corrections
in $1/L$ since no further powers of $1/\log L$  appear in
the expansion \cite{CalabreseRemark}. 

Similar conformal field theory manipulations were used to show
that the von Neumann entropy for the reduced density matrix
is given by 
$S(\rho_L)=-(c/6) \log \tilde q+ O\left(1/L\right)$ \cite{Holzhey},
which implies a direct relation between
entropy and single-copy entanglement
\be
\label{relation}
E_1(\rho_L)=\frac{1}{2} S(\rho_L)- \frac{c}{6} \frac{\pi^2}{\log L}+
O((1/L)\log{L}), 
\ee
the last subleading 
correction being easily calculated from the results in Ref.\
\cite{Holzhey}. 
This result fixes completely the 
value of the leading eigenvalue of the reduced density 
matrix of the block of size $L$ to be dictated by its
entropy, that is, 
$\lim_{L\to \infty}  \log (1/\lambda_1)/S(\rho_L)=1/2$.
Corrections to this limit can be obtained from Eq.\
(\ref{relation}).
Quite remarkably, all the eigenvalues will inherit the same leading behavior
and differ by their subleading corrections controlled by the 
positive eigenvalues of $(L_0+\bar L_0)$.  This result
establishes the quantitative connection between the 
single-copy entanglement and the geometric entropy in all
critical systems that can be described in the framework of
conformal field theory.

{\it Spin chains corresponding to general quasi-free
fermionic models. --} We will aim at 
strenghening the previously 
achieved result by investigating the same question in a 
different setting: we will
 investigate all translationally invariant 
spin models that can, under 
a Jordan-Wigner transformation,
be written as an isotropic 
quadratic Hamiltonian in fermionic operators. 
This setting includes the XX model.

The Jordan-Wigner transformation relates the Pauli operators
in the spin system to fermionic operators 
obeying $\{  c_j,   c_k\}=0$ and
$\{  c_j^\dagger,   c_k\}=\delta_{j,k}$, according to
\begin{equation}
	{\sigma}_l^x=\prod_{k<l}{\sigma}_k^z({c}_l+{c}_l^	\dagger),\,
      i{\sigma}_l^y=\prod_{k<l}{\sigma}_k^z({c}_l-
      \hat{c}_l^	\dagger),\,
	{\sigma}_l^z=1-2{c}_l^\dagger{c}_l.
\end{equation}
%
%\begin{eqnarray}
%	  \sigma_l^x &=& 
%	\frac{1}{2} \prod_{n=1}^{l-1} (1- 2   c_n^\dagger   c_n)
%	(  c_l^\dagger +   c_l),\\
%	  \sigma_l^y &=& \frac{1}{2i} \prod_{n=1}^{l-1} 
%	(  c_l^\dagger -   c_l)
%	(1- 2   c_n^\dagger   c_n),\,\,
%	  \sigma_l^z =   c_l^\dagger   c_l -\frac{1}{2}.\,\,
%\end{eqnarray}
The ground state is a quasi-free fermionic state, so a 
state that is
completely characterized by the second moments 
of fermionic operators.
Consider now such an infinite spin chain that 
corresponds to a general translationally invariant 
isotropic quasi-free fermionic model. These 
embody chain systems the Hamiltonian of which can be
cast into the form $H= \sum_{l,k} c_l^\dagger A_{ l-k } c_k$,	
with some general $A_l=A_{-l}\in\rr$ of which we do not
make an assumption. 
%This class of spin chains includes
%models that do not correspond to a conformal field theory.
%
The statement we arrive at is the following: 
if the entropy satisfies
\begin{equation}\label{assumption}
	S(\rho_L) = \xi \log L  + O(1),	
\end{equation}
for some $\xi>0$, 
then the single-copy entanglement satisfies
\begin{equation}\label{result}
	E_1(\rho_L)  = \frac{1}{2} S(\rho_L) + O(1).
\end{equation}
That is, if we find that the entropy of entanglement
scales asymptotically as the logarithm of $L$ -- as encountered
in this class of systems exactly at criticality --
then we can infer that the single-copy entanglement will
be asymptotically exactly one half of it, in the leading
order terms. 
This does notably not fix such a relationship
in case that, for example, the system is gapped and 
the entropy of entanglement saturates.
This statement follows from the subsequent argument.

\begin{figure}
\psfrag{a}{\small $i \delta$}
\psfrag{b}{\small $-i \delta$}
\psfrag{c}{\small $\epsilon/2$}
\psfrag{d}{\small $1+\epsilon$}
\psfrag{e}{\small $-1-\epsilon$}
\centering
\includegraphics[width=.42\textwidth]{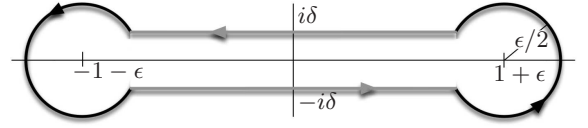}
\caption{Contour of integration to be taken in case of both
the entropy of entanglement and the single-copy 
entanglement.}\label{fig1}
\end{figure}

The reduced state of a block of length $L$
is entirely specified by the eigenvalues of the
real symmetric 
$L\times L$  Toeplitz matrix $T_L$, which 
defines the second moments of fermionic operators.
 The fact that $T_L$ is a 
Toeplitz matrix reflects the translational invariance of the model,
being symmetic 
follows from the isotropy. 
The $l$-th row of this matrix is given by 
	$(t_{-l+1},t_{-l+2}, ..., t_0 ,...,t_{L-l})$
\cite{numbers}.
The latter represents 
the Fermi surface, 
and essentially characterizes the fermionic model.
The eigenvalues of $T_L$
will be labeled as $\mu_1,...\mu_L \in[-1,1]$, which are
the zeros of the characteristic polynomial
$F:\cc\rightarrow \cc$,
%\begin{equation}
	$F(z )= \det[ z \id_L - T_L ]$. 
%\end{equation}
This function $F$ is meromorphic, and all its 
real zeros are  contained in
the interval $[-1,1]$, corresponding to the spectrum of $T_L$. 
The entropy of entanglement can be obtained as 
$S(\rho_L) = \sum_{l=1}^{L} f_S(1,\mu_l)$
\cite{Korepin1,Korepin2},
where $f_S:\rr^+\times \cc\rightarrow \cc$ as a complex embedding 
is defined as 
$f_S(x,y) = -((x+y)/2)\log ((x+y)/2) - ((x-y)/2)\log
((x-y)/2)$
%$f_S(x,y) = -\frac{x+y}{2}\log \frac{x+y}{2} - \frac{x-y}{2}\log
%\frac{x-y}{2}$. This function extends to the complex plane, 
to avoid problems with non-analyticities.
Actually, we can write \cite{Korepin1,Korepin2}
\begin{equation}
	S(\rho_L)= \lim_{\varepsilon \searrow 0 }
	\lim_{\delta\searrow 0 }
	\frac{1}{2\pi i}
	\int f_S(1+\varepsilon, z) \frac{F'(z)}{F(z)} dz.
\end{equation}
The contour of the integration is shown in Fig.\ \ref{fig1},
which is as in Ref.\ \cite{Singleshot}, but slightly different
from the one in Ref.\ \cite{Korepin1}. 
In turn, we may write 
$-\log \lambda_1 = 
\sum_{l=1}^{L} f_1(0,\mu_l)$ 
\cite{Singleshot},
in terms of the above $\mu_1,...,\mu_L$, where now 
$f_1:\rr^+ \times \cc \rightarrow\cc$, $f_1 (\varepsilon,z) = -\log ( (1+  
	(z^2+\varepsilon^2)^{1/2})/2)$. Respecting the
cuts of the logarithm (see Refs.\ \cite{Singleshot} and
\cite{Footnote}), we may cast $-\log \lambda_1$
and hence (up to integer brackets)
$E_1(\rho_L)$ into the form
\begin{equation}
	- \log\lambda_1 =\lim_{\varepsilon \searrow 0 } 
	\lim_{\delta \searrow 0 }
	\frac{1}{2\pi i}
	\int f_1(\varepsilon, z) \frac{F'(z)}{F(z)} dz.
\end{equation}
Now we know that $T_L$ is a real symmetric Toeplitz matrix,
which means that we can assess the asymptotic behavior
of their determinants. This can be done using proven
instances of the Fisher-Hartwig
conjecture \cite{Korepin2,Lieb}; proven instances, as 
we consider isotropic models \cite{Korepin2}. 
Concerning the function $F:\cc\rightarrow \cc$, 
this observation enables us to write 
\begin{equation}
\frac{F'(z)}{F(z)}
	 =a (z)  L -  b(z)  \log L   + O(1),
\end{equation}
where $b(z)= -2 R  \beta(z)\beta'(z)$,
with $\beta:\cc\rightarrow\cc$ being a function defined as
$\beta(z)= \log((z+1)/(z-1))/(2 \pi i)$, see Refs.\ \cite{Korepin2}. 
$R$ in turn 
is half
the number of discontinuities of  the above symbol in 
the interval $[0,2\pi)$. For the XX model,
e.g., we have that $R=1$.
Now, if Eq.\ (\ref{assumption}) is valid, 
then
\begin{equation}
	\lim_{\varepsilon \searrow 0 }
	\lim_{\delta\searrow 0 }
	\int f_S(1+\varepsilon, z) a(z) dz=0.
\end{equation}
But since $S(\rho_L)\geq E_1(\rho_L)$ for all $L\in \nn$,
necessarily
	$\lim_{\varepsilon \searrow 0 } 
	\lim_{\delta \searrow 0 }
	\int f_1(\varepsilon, z) a(z) dz=0$
must hold. Hence, we only have to consider the 
logarithmically divergent term.
It is sufficient for our argument, therefore, 
for the entropy of entanglement and 
for the single-copy entanglement
to consider the contour integrals
\begin{eqnarray}
	I_S &=& \lim_{\varepsilon \searrow 0 }
	\lim_{\delta\searrow 0 }
	\frac{1}{2\pi i}
	\int f_S(1+\varepsilon, z) b(z) dz,\\
	I_1 &=& \lim_{\varepsilon \searrow 0 }
	\lim_{\delta\searrow 0 }
	\frac{1}{2\pi i}
	\int f_1(\varepsilon, z) b(z) dz.
\end{eqnarray}
$b$ is analytic outside $[-1,1]$. 
In turn, this means that the contributions of the circle pieces vanish in both
cases.
Hence, we finally arrive at 
\begin{eqnarray}
      S(\rho_L) &=& \frac{R}{\pi^2} 
      \int_{-1}^1 dx
	\frac{f_S(1,x)}{1-x^2} \log L + O(1),\\
	E_1(\rho_L) &=&  \frac{R}{\pi^2}  \int_{-1}^1 dx
	\frac{f_1(0,x)}{1-x^2} \log L  + O(1).
\end{eqnarray}
Since $f_1(0,x)=- \log (1+|x|)/2$ for $x\in[-1,1]$, this gives 
\begin{equation}
	S= \frac{R}{3}\log L + O(1),\,\,\,
	 E_1 = \frac{R}{6}\log L + O(1),
\end{equation}
which in turn implies the validity of Eq.\ (\ref{result}): 
In these models,
whenever the system is critical, the single-copy entanglement is 
again exactly half the asymptotically available 
in its leading contribution. This gives further substance to the
previous consideration in a language of 
conformal field theory.

%\begin{figure}
%\centering 
%\includegraphics[width=.50\textwidth]{NewFigure.ps}
%\caption{Contour of integration to be taken in case of both
%the entropy of entanglement and the single-copy entanglement.}
%\label{fig1}
%\end{figure}

{\it Single-copy entanglement away from criticality. --} The relation 
between single-copy entanglement
and entropy can be demonstrated 
near critical points in
some integrable models. We illustrate this fact
using the XY spin chain, in a slightly different set-up: we
consider the chain of length $N$ with 
periodic boundary conditions, 
where the half chain $L=N/2$ constitutes one system. For
large $N$, the density matrix of the system can be
arbitrarily well approximated in trace-norm by
\be
	\frac{e^{-H}}{{\text{tr}}[e^{-H}]}, \,\,
	H=\sum_{k} \epsilon_k d^\dagger_k d_k,\,\,
\epsilon_{k} = 
\begin{cases}
2 k  \epsilon \ , & {\rm if} \ \lambda < 1 \\
(2 k+1)  \epsilon \ , & {\rm if} \ \lambda > 1 \ , 
\end{cases}
\label{dispersion}
\ee
\cite{Peschel}.
Here, $k \in \nn$, $\lambda\in \rr$ 
is the parameter controlling the external magnetic field, 
$\lambda^*=1$ corresponds to the quantum phase transition point, 
and
$ \epsilon = \pi (I(({1-x^2})^{1/2}))/{I(x)} $, $I:\cc\rightarrow\cc$ 
is the complete 
elliptic integral of the first kind, $I(x) = \int_0^{\pi/2} d
  \theta/(1 - x^2 \sin^2 (\theta))^{1/2}$.
$x$ is related to $\lambda$ and $\gamma$: 
\be
x = 
\begin{cases}
({\lambda^2 + \gamma^2 - 1})^{1/2}/\gamma \ , 
	& {\rm if} \ \lambda < 1, \\
\gamma / ({\lambda^2 + \gamma^2 - 1})^{1/2}  \ , 
	& {\rm if} \ \lambda > 1 \ , 
\end{cases}
\label{eks}
\ee
with the condition $\lambda^2 + \gamma^2 > 1$ (external region of the BM-circle
\cite{Lieb}). A computation of the single-copy entanglement  
with respect to this partitioning 
can be performed in terms of $\epsilon$, transforming sums into 
integrals by means of the Euler-McLaurin expansion, and finding
\be 
\label{singlecopynoncritical1}
%E_1(\rho_{L,\epsilon})
-\log\lambda_1
= \left\{
\begin{array}{ll}
\frac{\pi^2}{24 \epsilon}- \frac{\epsilon}{24}
+ O(e^{-\epsilon})& \text{ if } \lambda <1,\\
%\label{singlecopynoncritical2}
\frac{\pi^2}{24 \epsilon}+\frac{\log 2}{2}+ \frac{\epsilon}{12}
+O(e^{-\epsilon}) &  \text{ if }  \lambda >1.
\end{array}\right.
\ee
No 
subleading corrections in powers of $\epsilon$ do appear
in the expansion. On the other hand it can be seen by explicit evaluation that
that the entropy of the reduction $\rho_{L,\epsilon}$
can be related in this case to the single copy-entanglement by $
S(\rho_{L,\epsilon})
=- \left(1-\epsilon\frac{\partial}{\partial \epsilon}\right)
\log\lambda_1$, which shows that
\be
\label{relationnoncritical}
	\lim_{\epsilon\rightarrow0} 
	\frac{E_1(\rho_{L,\epsilon})}{S(\rho_{L,\epsilon})}
	=\frac{1}{2}.
\ee
This is precisely the limit where
the theory becomes critical.

\vspace{3pt}

In this work we have shown that the leading critical scaling of the
single-copy entanglement is exactly one half of the entropy of entanglement in
critical quantum spin chains, using tools of conformal field theory. We have
also provided an analysis for all translationally-invariant quantum
spin chains that can be mapped onto an isotropic quasi-free fermionic model
under a Jordan-Wigner transformation, leading to similar conclusions. Away
from criticality, this simple relation is recovered when approaching the
quantum phase transition point, as seen in the XY model. It is a fact that 
the single-copy entanglement could be experimentally studied in, for instance, 
systems of cold atoms in optical lattices, ions in ion-traps, or solid-state
devices. Our hope is that the results we have presented here will
serve as guideline for that kind of experiments, as well as for a better
understanding of the structure of the ground state correlations in quantum spin chains.   

\vspace{3pt}

{\it Acknowledgements. --} Two of us (JE and JIL) would like to 
thank the members of the
Perimeter Institute for Theoretical Physics
-- where part of the work was
carried out --  in particular D.\ Gottesman and M. Mosca 
for kind hospitality. We would like to thank 
P.\ Calabrese,
J.I.\ Cirac,
V.E.\ Korepin,
T.\ Osborne,
M.B.\ Plenio,
R.F.\ Werner, and
M.M.\ Wolf for 
discussions.
This work has been supported by the DFG
(SPP 1116, SPP 1078),
the EU (QUPRODIS, QAP), the EPSRC,  the
EURYI scheme, and the MEC (Spain).

{\it Note added:} After completion of this work, we 
became
aware of the independent 
work Ref.\ \cite{PeschelNew}, where
the first (leading-order) term for single-copy entanglement
in the conformal case was also
discussed in detail and clarity.


\begin{thebibliography}{99}
\bibitem{Basis}
	Note that $|i  , j\rangle$, $i,j=1,...,M$, are elements 
	of a product basis with respect to the distinguished block
	  and the rest of the chain.
\bibitem{Prob}	 	       	
	In a probabilistic version $E_p$,
	one may also allow for maximally entangled states of
	different dimensions with certain probabilities, 
	compare Ref.\ \cite{Dan}.
\bibitem{loga}
	We define the entropy, as well as the single-copy 		entanglement,	
	in terms of the natural logarithm.
\bibitem{cutoff}
This value of the cut-off is to be understood in all the 
forthcoming 
calculations.
\bibitem{modular}
	The needed transformation 
	is $\tau \mapsto -1/\tau$,
	which amounts to 
	$Z_L(q)=Z_L(\tilde q)$, with $q=e^{-2 \pi^2/\log L}$
	and 	$\tilde q=e^{-2\log L}$.
\bibitem{CalabreseRemark}
	Note that the leading order term also follows from the
	results on the expressions for the Renyi 
	   entropies \cite{Private}
	in the conformal setting in Ref.\ \cite{Cardy}. The non-leading term can be explicitely
checked by means of numerical evaluations or analytical derivations for
integrable models like the one in Ref.\ 
\cite{Peschelchain}.
\bibitem{numbers}
	These numbers are for an infinite chain found to be 
	$t_l =   \int_0^{2\pi} g(k) e^{-ilk} dk/(2 \pi)$,
	where $g:\cc\rightarrow \cc$ is the symbol
	\cite{Lieb}.
\end{thebibliography}
\end{document}